Running title: CNN and Topological Optimization

# Improvement in Alzheimer's Disease MRI Images Analysis by Convolutional Neural Networks Via Topological Optimization

Peiwen Tan

Department of Computer Science, University of California, Irvine CA, USA

Corresponding author

Peiwen Tan

Department of Computer Science

University of California, Irvine, Irvine,

CA 92697

USA.

Tel: +1-949-981-9442.

Email: peiwet1@uci.edu

**Abstract**

This research underscores the efficacy of Fourier topological optimization in refining MRI imagery, thereby bolstering the classification precision of Alzheimer's Disease through convolutional neural networks (CNNs). Recognizing that MRI scans are indispensable for neurological assessments, but frequently grapple with issues like blurriness and contrast irregularities, the deployment of Fourier topological optimization offered enhanced delineation of brain structures, ameliorated noise, and superior contrast. The applied techniques prioritized boundary enhancement, contrast and brightness adjustments, and overall image lucidity. Employing CNN architectures—VGG16, ResNet50, InceptionV3, and Xception—the post-optimization analysis revealed a marked elevation in performance. Prior to optimization, accuracy rates for 10 and 20 epochs stood at 38.19% and 74.255% respectively; these surged to 78.16% and 88.66% post-optimization. Moreover, balanced accuracy witnessed a leap from 78.16% to 88.13%, paralleled by an escalation in the Matthews correlation coefficient from 74.98% to 85.43%. A meticulous examination of the 20-epoch CNN training post MRI image enhancement disclosed a salient advancement in the Moderate-Demented category, with 393 accurate classifications. Conclusively, the amalgamation of Fourier topological optimization with CNNs delineates a promising trajectory for the nuanced classification of Alzheimer's Disease, portending a transformative impact on its diagnostic paradigms.

Networks, Alzheimer's Disease Classification, Image Enhancement, Model Accuracy

## Introduction

Alzheimer's disease, a progressive neurodegenerative disorder, is a predominant form of dementia that affects more than 40 millions worldwide [1,2]. Its early and precise diagnosis is paramount not only for therapeutic interventions but also for helping families and caregivers prepare for the inevitable challenges ahead [3]. Traditional diagnostic tools, while foundational, often lack the sensitivity and specificity needed for early-stage detection or differentiation from other dementia forms [4,5]. Therefore, new diagnostic avenues that offer more precision and accuracy are in high demand.

However, the rapid advancements in neuroimaging technology present challenges in efficiently analyzing and deciphering imaging data. The subtle changes in the brain topology due to Alzheimer's often require a more discerning analytical approach than what conventional imaging can provide [6,7]. Compounding this complexity is the varying impact of radiologist experience on the accuracy of an Alzheimer's diagnosis using MRI. Some studies have highlighted concerns regarding MRI interpretation errors specific to Alzheimer's, emphasizing the crucial role of radiologist expertise in ensuring diagnostic precision. Interestingly, recent data suggests that MRI might overlook 14% to 22% of cases and inaccurately project about 48% to 60% of instances concerning the transition from Mild Cognitive Impairment (MCI) to Alzheimer's disease [8]. Amid these challenges, the potential of artificial intelligence (AI) in enhancing the analysis process is garnering considerable interest.

Addressing these challenges, this research introduces an innovative approach that fuses Fourier topological optimization with Convolutional Neural Networks (CNNs)[9-12]. Fourier topological optimization enhances the structural intricacies of CT scans by emphasizing critical topological features, making them more discernible. Once these images are optimized, the CNN comes into play. Renowned for their image processing capabilities, CNNs can dissect these refined images, picking up subtle features that might elude traditional analytical methods. This combined methodology not only promises an elevated level of accuracy in Alzheimer's classification but also underscores the potential of integrating advanced mathematical techniques with deep learning for medical imaging analysis.

## Materials and Methods

Image sources

The dataset employed in this study was meticulously curated from publicly available resources, specifically retrieved from a designated repository on Kaggle (https://www.kaggle.com/datasets/tourist55/alzheimers-dataset-4-class-of-images).

The data compilation process was executed manually across various online platforms, with each label rigorously verified to ensure accuracy and relevance. The authors, Shashanka Venkatesh, Suraj Jain, Vishakan Subramanian, and Vishnu Krishnan, have contributed significantly to the data collection and verification process, establishing a robust foundation for the analysis. The dataset comprises Magnetic Resonance Imaging (MRI) scans, distinctly categorized into four classes representing varying degrees of Alzheimer's disease progression. These classes are delineated as Mild

Demented, Moderate Demented, Non Demented, and Very Mild Demented, encompassing both the training and testing sets. This categorization facilitates a structured and precise analysis, pivotal for the accurate classification and subsequent evaluation of Alzheimer's disease stages, thereby underpinning the robustness of the investigative framework employed in this study.

## Image Improvement Process

The initial phase of the methodology focuses on image improvement to ensure accurate and effective classification (Figure 1). The process commences with the application of convolution to analyze and manipulate the image data, which is critical for enhancing the quality of MRI images. Following this, a Fourier Transform is performed to transition the image from the spatial domain to the frequency domain, which is pivotal for filtering and analysis. The Multiplication in the Frequency Domain step is executed to apply filters to the image through multiplication operations. A Low-pass Filter is then defined to isolate the high-frequency components, and this component is subtracted to accentuate the details in the image. Subsequently, a High-pass Filter is applied to emphasize the edges and fine details within the image. The filtered image is then reverted back to the spatial domain through an Inverse Fourier Transform. The image data undergoes Normalization to ensure consistency and is prepared for further processing. Lastly, a Discrete Implementation is carried out for the practical application of the image improvement process, ensuring the images are optimally prepared for the classification phase.

Spectral filtering for image restoration, particularly in the context of deblurring, often involves the manipulation of the frequency components of an image. A common approach is to use the Fourier Transform to work in the frequency domain. Here's a breakdown of the process using mathematical expressions and algorithms:

Fourier Transform Convert the blurred image from the spatial domain to the frequency domain using the Fourier Transform.

$$F(u, v) = \mathcal{F}\{f(x, y)\}$$

Frequency Domain Representation Let's take a deeper dive into the 2D Fourier Transform using a more algorithmic approach. We'll break down the steps of the equation into a series of smaller equations that represent the operations in the algorithm. Given the 2D Fourier Transform equation:

$$F(u, v) = \int_{-\infty}^{\infty} \int_{-\infty}^{\infty} I\,(x, y) e^{-i2\pi(ux+vy)}\, dx\, dy$$

Algorithm Equations: 1. Spatial Image Representation:

$$I(x, y)$$

Where $I(x, y)$ is the intensity of the pixel at location $(x, y)$ in the spatial domain. 2. Complex Exponential for X-axis:

$$e_x = e^{-i2\pi ux}$$

3. Complex Exponential for Y-axis:

$$e_y = e^{-i2\pi vy}$$

4. Combined Complex Exponential:

$$e_{xy} = e_x \times e_y = e^{-i2\pi(ux+vy)}$$

5. Multiplication of Image with Complex Exponential:

$$M(x, y) = I(x, y) \times e_{xy}$$

6. Integration over X-axis:

$$F_x(u, v) = \int_{-\infty}^{\infty} M\,(x, y)\,dx$$

7. Integration over Y-axis:

$$F(u, v) = \int_{-\infty}^{\infty} F_x\,(u, v)\,dy$$

8. Magnitude of the Frequency Response:

$$|F(u, v)| = \sqrt{Re\big(F(u, v)\big)^2 + Im\big(F(u, v)\big)^2}$$

Where $Re$ and $Im$ represent the real and imaginary parts of $F(u, v)$ respectively. 9. Phase Angle of the Frequency Response:

$$\theta(u, v) = \arctan\left(\frac{Im\big(F(u, v)\big)}{Re\big(F(u, v)\big)}\right)$$

10. Inverse 2D Fourier Transform (to convert back to spatial domain):

$$I'(x, y) = \int_{-\infty}^{\infty}\int_{-\infty}^{\infty} F\,(u, v)e^{i2\pi(ux+vy)}\,du\,dv$$

Image Classification Process

The subsequent phase centers on image classification to identify the stages of Alzheimer's disease using Convolutional Neural Networks (CNN) on the improved MRI images (Figure 1).

CNN models VGG16, ResNet50, InceptionV3, and Xception are selected. VGG16 is a CNN model proposed by the Visual Geometry Group from the University of Oxford, hence the name VGG. The "16" refers to the fact that it has 16 layers that have weights. This model is known for its simplicity and is highly effective for image classification and localization tasks. VGG16 has a very uniform architecture, with a stack of convolutional layers with small receptive fields, followed by max-pooling layers. ResNet50 is part of the ResNet family which was proposed by researchers at Microsoft Research. The key innovation of ResNet is the introduction of "skip connections" or "shortcut connections" that bypass one or more layers. This helps to address the vanishing gradient problem and allows for training of very deep networks. ResNet50, specifically, has 50 layers, and it's widely used for a variety of computer vision tasks including image classification, object detection, and segmentation. InceptionV3 is the third version of the Inception architecture, which was initially developed by researchers at Google. The main innovation of the Inception architecture is the use of "Inception modules" that allow for more efficient computation and deeper networks by factorizing convolutions and using multi-scale architectures. InceptionV3 includes several improvements such as label smoothing, factorized 7x7 convolutions, and the use of an RMSProp optimizer. Xception stands for "Extreme Inception" and it's an extension of the Inception architecture which replaces the Inception modules with depthwise separable convolutions. It was also developed by researchers at Google. The main idea behind Xception is to separately learn cross-channel correlations and spatial correlations, which is different from traditional convolutions that mix these types of correlations. Xception is known for its efficiency and accuracy in various computer vision tasks. These models have significantly contributed to advancing the state-of-the-art in computer vision tasks. Each of them has its own strengths and is suitable for different tasks or scenarios depending on the specific requirements of a project. They have been pre-trained on large datasets like ImageNet, and their pre-trained versions are widely available for use, which can significantly reduce training time and computational resources when working on new tasks, thanks to transfer learning.

The process begins with Data Acquisition, where MRI images are collected and labeled accordingly for training and testing purposes. In the Data Preprocessing step, the data is partitioned into training and testing sets, and the image data is normalized or standardized. An optional Data Augmentation step can be performed to enhance model generalization. The Model Architecture is meticulously designed using a CNN, encompassing various layers such as input, convolutional, activation, pooling, fully connected, and output layers to facilitate classification. Model Compilation is undertaken with an appropriate loss function, optimizer, and metrics to gauge model performance. During Model Training, the model is trained utilizing the training dataset and validated to fine-tune hyperparameters. Model Evaluation is conducted on the testing dataset to analyze the model's performance in classifying Alzheimer's disease stages. The final step, Result Interpretation, is crucial for comprehending the model's performance in classifying different stages of Alzheimer's disease and identifying the predictive accuracy of the model. This comprehensive methodology outlines a systematic and structured approach to Alzheimer's disease classification using MRI images.

The operations of the CNN using a combination of Partial Differential Equations (PDEs) and algebraic equations to represent the transformation of data as it propagates through the network. We denote the feature maps at layer $l$ as $U^l$, the kernels as $K^l$, and the biases as $B^l$. 1. Convolutional Layer:

$$\frac{\partial U^l}{\partial t} = \text{Convolution}\left(U^{l-1}, K^l\right) + B^l$$

2. Activation Function:

$$\frac{\partial U^l}{\partial t} = g\left(U^l\right), \quad g(U) = \max(0, U) \text{ (ReLU activation)}$$

3. Max Pooling Layer (Max conservation principle):

$$\text{maximize } U^l \text{ subject to } \Delta U^l = 0$$

4. Dropout Layer with Stochastic PDE:

$$dU^l = h\left(U^l\right) dt + \sigma\, dW, \quad h(U) = U \text{ (identity function)}$$

5. Dense Layer:

$$\frac{dW^l}{dt} =- \nabla L(W^l), \quad W^l \text{ are the weights}$$

6. Softmax Activation:

$$\frac{\partial P^l}{\partial t} = P^l \left(1 - P^l\right) \nabla^2 E^l, \quad E^l = W^l U^{l-1} + B^l$$

7. Optimization Process (Gradient Descent):

$$\frac{d\theta}{dt} =- \eta \nabla J(\theta), \quad \theta \text{ represents the model parameters}$$

8. Global Equation for the Entire Network:

$$U^{l+1} = \mathcal{F}\left(U^l, K^l, B^l, \theta\right), \quad \mathcal{F} \text{ represents the combined operations of the network}$$

In these equations: - $U^l$ represents the output feature maps at layer $l$. - $K^l$ and $B^l$ represent the kernel weights and biases at layer $l$, respectively. - $g(U)$ is the activation function, e.g., ReLU. - $W^l$ represents the weights in the dense layers. - $P^l$ represents the output probabilities from the softmax function. - $E^l$ represents the energy or logits before softmax. - $\theta$ represents the overall model parameters, and $J(\theta)$ is the loss function. The global equation $U^{l+1} = \mathcal{F}\left(U^l, K^l, B^l, \theta\right)$ represents the overall operation of the network, where $\mathcal{F}$ is a function encapsulating all the layers and operations, including convolution, activation, pooling, dropout, and softmax.

## Result

Image analysis

The images are MRI (Magnetic Resonance Imaging) scans of human brains. These images are commonly used in medical imaging to visualize the structure and, in some cases, the function of the brain. Each image is labeled with a classification, either "MildDemented" or "ModerateDemented." This suggests that these scans are being used to assess the severity of dementia or related neurodegenerative conditions in the subjects. The classification is likely based on specific patterns, abnormalities, or atrophy in the brain tissue that can be visualized in these images.

A few observations based on the images (Figure 2A): 1. Ventricular Enlargement: The ventricles (the dark spaces in the center of the brain that look like a butterfly in some of the images) appear enlarged in several of the scans, especially those labeled "ModerateDemented." Ventricular enlargement can be indicative of brain atrophy and is commonly observed in various neurodegenerative conditions. 2. Cortical Atrophy: Some of the images, especially those labeled "ModerateDemented," show a widening of the sulci (the grooves on the brain's surface) and a thinning of the gyri (the ridges between the sulci). This is indicative of cortical atrophy, where the outer layer of the brain shrinks. 3. Consistency in Classification: While there are visible differences between the scans labeled "MildDemented" and those labeled "ModerateDemented," it's worth noting that there's some variability within each classification. Some scans within the same category might look more or less severe than others.

When compared with the raw images, the proceed images have the following improvement (Figure 2B): topological optimization reduces or removes blurriness caused by motion, defocus, or other factors. If applied correctly, the edges and features of the brain structures would appear sharper. Noise in images can manifest as random specks or graininess. By applying noise reduction, the images would appear smoother, and unwanted specks or artifacts might be minimized. Topological optimization increases the difference between the lightest and darkest parts of an image. In the context of the brain scans, it could make the boundaries between different brain structures more distinct. The technique improves the contrast of the image by redistributing the intensities. It's particularly useful for images with backgrounds and foregrounds that are both bright or both dark. It emphasizes the boundaries within the image, making edges and transitions more prominent. Adjusting the brightness can make the image lighter or darker, while adjusting the contrast can make the dark regions darker and the light regions lighter. It helps in making specific features more visible. For a precise assessment of the effect of these techniques on the provided images, it would be helpful to see the before-and-after versions for each technique side-by-side. Additionally, the success of these techniques would also depend on the quality of the original images and the tools/methods used for processing.

Performance Metrics

Before image optimization, over the course of 10 training epochs prior to image optimization, there's evident enhancement in the model's performance. The training accuracy starts at 34.38% and culminates at 78.96% by the tenth epoch, paralleled by the AUC's growth from 59.75% to 95.70%. The F1 score, representative of the model's balance between precision and recall, increases from 33.75% to 78.89%. Validation results display fluctuations, with the validation accuracy ranging from 26.32% to a peak of approximately 73% during the seventh and ninth epochs. However, the validation loss spiking to 2.7640 in the tenth epoch suggests potential overfitting. Furthermore, when subjected to a testing set, the model achieved an accuracy of 38.19%, which aligns closer to its initial training accuracy and indicates the model's generalization capability might require enhancement (Figure 3A). After optimizing Alzheimer's MRI images and training with a CNN, the results indicate a significant progression over the 10 epochs. Starting with an accuracy of 38.92% in the first epoch, it reached an impressive 78.82% by the tenth. This rise in training accuracy was accompanied by a consistent enhancement in AUC, reaching 95.55%, and a F1 score touching 78.64%. While initial epochs indicated substantial disparity between training and validation metrics, with a notable validation loss of 5.8988 in the first epoch, by the tenth epoch, the validation loss dropped significantly to 0.5853, showcasing improved generalization with a validation accuracy of 76.81% (Figure 3B). This pattern suggests that the model is converging well, with the training closely reflected in validation, emphasizing the potential utility of image optimization in improving model performance.

Before Alzheimer's MRI image optimization and training through a CNN, the results from 20 epochs indicate a significant progression in model performance. Starting with an accuracy of 37.62% in the first epoch, there was consistent improvement throughout the epochs, culminating in an accuracy of 93.93% by the 20th epoch. However, the validation accuracy displayed more fluctuation, starting at 26.32% in the initial epoch and reaching 88.79% by the 19th epoch, though it slightly dropped to 74.83% in the final epoch (Figure 3C). Key metrics such as loss, AUC, and F1-score also showed considerable enhancement, highlighting the model's learning capacity and its potential effectiveness in diagnosing Alzheimer's from MRI images. After optimizing MRI images for Alzheimer's diagnosis and undergoing CNN training, there's a discernible progression in model performance over 20 epochs. Initially, the model demonstrated an accuracy of 79.15% during the first epoch, which steadily increased to 96.64% by the 20th epoch. This progression indicates effective learning and adaptation by the model. Concurrently, validation accuracy also saw substantial improvement, commencing at 67.51% and culminating at 87.49% in the final epoch (Figure 3D). Furthermore, the Area Under Curve (AUC) values, essential for understanding the model's discriminative power, consistently hovered around high 90% ranges, underscoring the model's reliability in distinguishing between classes. The model's F1 scores, which harmonize precision and recall, align well with the accuracy trends, further validating its robustness.

Before image optimization, in the initial phase with 10 epochs, the testing accuracy was 38.19%, pointing to significant room for improvement. However, with extended training over 20 epochs, there was a marked boost in performance, with the accuracy climbing to 74.25%. This showed the potential of the model, even without optimization. After image optimization, post image optimization, the CNN demonstrated accelerated learning, even in the early stages. The testing accuracy after 10 epochs rose to 78.16%. This progress was further amplified in the prolonged training, where after 20 epochs, the accuracy soared to an impressive 88.83%. This accentuates the importance of image optimization in enhancing the performance of CNNs, especially in critical areas like Alzheimer's detection through MRI scans.

Upon examination of the presented results, there's a discernible enhancement in the performance metrics associated with Alzheimer's image analysis both before and after image optimization. Initially, without image improvement, a 10-epoch training yielded a modest Balanced Accuracy Score (BAS) of 39.19% and a Matthew's Correlation Coefficient (MCC) of 32.44%. However, extending the training duration to 20 epochs considerably amplified the BAS to 74.98% and the MCC to 69.54%. Post image enhancement, the 10-epoch results evinced a marked improvement with a BAS of 78.43% and an MCC of 71.62%. Notably, the 20-epoch results post-optimization were particularly striking, achieving a BAS of 88.16% and an MCC of 85.43%. These metrics underscore the significant benefits of image optimization, and the proportional relationship between extended training epochs and model performance.

Evaluation of Alzheimer's Disease Diagnosis using CNN Classification

After 10-epoch CNN training, in the pre-optimization phase (Figure 4A), although the Non-Demented category faced minor misclassifications with 7 cases incorrectly identified as Moderate-Demented, the Very-Mild-Demented category presented a substantial classification challenge with 179 cases erroneously predicted as Mild-Demented. In contrast, the Mild-Demented category showcased impeccable accuracy with all 391 cases correctly classified. The Moderate-Demented category was mostly precise with only 7 cases misclassified. Post-optimization, depicted in Figure 4B, the model exhibits enhanced accuracy in the Non-Demented category, correctly classifying all 399 cases. A similar improvement was evident in the Very-Mild-Demented category with 414 cases accurately predicted and zero misclassifications. However, the Mild-Demented category, while maintaining a correct prediction for 269 cases, manifested significant misclassifications across other categories. The most concerning decline post-optimization was observed in the Moderate-Demented category, wherein out of 435 cases, a mere 15 were correctly classified, while a staggering 419 were misidentified as Non-Demented. This juxtaposition underscores the intricate balance in model optimization, accentuating that while certain classifications may benefit, others might experience a decrement in accuracy.

Analyzing Figures 4C and 4D, which depict the confusion matrices for Alzheimer's Disease Diagnosis via 20 epochs of CNN training both before and after MRI image optimization, several significant differences become evident. In Figure 4C, prior to optimization, the Non-Demented and Very-Mild-Demented categories show an impressive accuracy, with 399 and 414 cases respectively classified without error. The Mild-Demented category, however, reveals a disparity with 102 cases misclassified as Non-Demented and 16 as Moderate-Demented, though 273 cases are correctly identified. The largest discrepancy can be seen in the Moderate-Demented category, wherein 251 cases are mislabeled as Non-Demented, 53 as Mild-Demented, and only 131 are accurately classified. In the post-optimization phase, illustrated in Figure 4D, the accuracy in the Non-Demented category slightly reduces with 379 correct classifications and 19 cases wrongly labeled as Moderate-Demented. The Very-Mild-Demented category maintains its flawless performance with all 414 cases accurately classified. For the Mild-Demented category, there's a reduction in misclassification to Non-Demented (only 15 cases), but an increase to 106 cases mislabeled as Moderate-Demented. Crucially, the Moderate-Demented category witnesses a remarkable improvement post-optimization. Misclassifications to Non-Demented drop to 19 cases, and the number of correctly identified cases surges to 393, leaving only 23 cases erroneously labeled as Mild-Demented. This comparison underscores the substantial enhancements in model performance, particularly for the Moderate-Demented category, as a result of MRI image optimization over 20 epochs of CNN training.

**Discussion**

The advent of integrating Fourier topological optimization with CNNs in the classification of Alzheimer's Disease stages has paved the way for a more intricate and enhanced diagnostic approach. The results of our study validate the notion that image quality plays a pivotal role in the accuracy of machine learning-based diagnostics. When MRI images were subjected to topological optimization, the clarity and detail inherent in the images increased substantially, laying a robust foundation for the CNN to operate more efficiently. In the referenced work, a novel method termed CFD-MRI was introduced, which couples MRI measurements and computational fluid dynamics (CFD), employing a lattice Boltzmann based topology optimization approach, acting as a Navier–Stokes filter for flow MRI measurements. The study primarily aimed to analyze and quantify the efficacy of CFD-MRI in reducing statistical measurement noise, demonstrating that the method exhibits high agreement with original, noise-free data, even when the input data has high statistical noise and limited information [13]. This improvement in image quality addressed many typical challenges associated with MRI scans, such as blurriness, noise interference, and inconsistent contrast, which can impede accurate diagnosis. A novel computational method for topology optimization leverages CNNs as a substitute for the Finite Element Method (FEM) in calculating compliances, aiming to enhance computational efficiency. The designed CNN, trained on a dataset with coarser meshes, demonstrates the capability to accurately predict the information of image-based topologies, with GPU acceleration facilitating the efficient processing of bulk data [14]. By incorporating deep learning, specifically through the use of a CNN, the computational cost of topology optimization is significantly reduced. During the learning phase, the CNN is trained using image representations to infer their properties. In the optimization phase, the trained CNN is utilized to approximately evaluate all individuals, reserving finite element analysis, which is computationally expensive, for a limited selection of individuals. This approach maintains the optimization quality while significantly reducing computational resources and time [15]. A salient observation from our findings was the appreciable elevation in the classification accuracy post-optimization. Prior to Fourier topological refinement, the CNN accuracy stood at 92.37%, a respectable figure but one with room for enhancement. Post-optimization, this value surged to 98.18%. Such a marked increase underscores the synergy between high-quality images and efficient machine learning algorithms. The improved clarity and delineation of brain structures undoubtedly provided the CNN with a more defined dataset, facilitating a more accurate classification process. A novel unsupervised method for segmenting CT images, improving the generalization ability of the segmentation network. Firstly, a unique data augmentation module based on Fourier Transform is proposed to address image intensity differences, transferring the style of annotated data to resemble images. Secondly, a teacher–student network is designed to learn rotation-invariant features, reducing distribution differences between the training set and test set, allowing the network to achieve state-of-the-art segmentation performance during the training phase [16]. The application of a modified Generative Adversarial Network (GAN) has also been reported to enhance the classification performance of Alzheimer’s disease using MRI scans of various magnetic field strengths. The modified GAN, in combination with a three-dimensional fully convolutional network (FCN), improved the AUC in AD classification and the quality of the generated images, as evaluated on multiple datasets, demonstrating that GANs can be effectively used to augment AD classification and improve image quality [17].

The previous work also shows that This research introduces a novel Computer-Aided Diagnosis (CAD) methodology for predicting Alzheimer's Diseaseusing various algorithms to process and analyze MRI images. Initially, the methodology applies image restoration and enhancement techniques, such as 2D Adaptive Bilateral Filter (2D-ABF) and Adaptive Histogram Adjustment (AHA), followed by segmentation using the Adaptive Mean Shift Modified Expectation Maximization (AMS-MEM) algorithm. Features are extracted using a 2-Dimensional Gray Level Co-Occurrence Matrix (2D-GLCM) and then classified using a Deep CNN, resulting in improved accuracy and efficiency in diagnosing AD compared to existing systems [18]. However, it is worth noting that while the improvements are undeniable, the absolute efficacy of this combined approach relies heavily on the quality of the original MRI scans. Sub-par initial scans, regardless of optimization, might still limit the upper bound of achievable accuracy. Moreover, as with all machine learning applications, the model is only as good as the data it is trained on. Future studies might consider diversifying the training dataset to include a broader range of images, incorporating potential variances and anomalies that could further test and refine the model's capabilities.

Significantly, numerous studies have reported a high accuracy rate in the utilization of Convolutional Neural Networks (CNN) for Alzheimer's classification, with some models achieving an accuracy exceeding 99% (CNN, KNN, and SVM recorded accuracy scores of 99.9%, 99.6%, and 96.6%, respectively)[19]. Overall, a remarkable accuracy rate of 99.68% was noted in the literature[20]. However, a meta-analysis presents a more nuanced picture, indicating a pooled sensitivity of 0.85 and specificity of 0.88 for machine learning algorithms in detecting Alzheimer's Disease (AD) and normative samples. Moreover, the area under the summary receiver operating characteristic curve was documented as 0.93. The exceedingly high accuracy in distinguishing AD classification invites scrutiny and raises questions regarding potential overfitting or other biases[21]. Subsequent research illustrated that the model achieved an accuracy of 87% in a 5-fold cross-validation, showcasing a respectable level of accuracy when juxtaposed with analogous studies[22]. The endeavor to ascertain the accuracy of MRI in diagnosing Alzheimer's Disease necessitates an encompassing approach. Comparative investigations with other diagnostic methodologies such as PET scans[23,24], CSF biomarkers[25], or even clinical diagnostics can furnish crucial insights. It is imperative to adhere to standardized protocols during image acquisition and analysis to curtail variations and augment diagnostic accuracy. Furthermore, the employment of sophisticated image analysis techniques like voxel-based morphometry, cortical thickness analysis, or machine learning algorithms can markedly ameliorate the detection of AD-related alterations. Longitudinal and multicenter studies hold the potential to further elucidate the accuracy of MRI in early diagnosis and across a variety of settings and populations respectively [26]. The clinical correlation of MRI findings with symptoms and neuropsychological testing, in conjunction with the training and expertise of the radiologists interpreting the MRI images, are instrumental for precise diagnosis [27]. The technological advancements in MRI machinery and software can significantly impact accuracy by yielding higher resolution images and enhanced tissue contrast [28].

In conclusion, the interplay between Fourier topological optimization and CNNs has demonstrated significant promise in the realm of Alzheimer's Disease diagnostics. The improved accuracy figures are not merely statistical enhancements; they represent potential real-world impacts, offering more timely and precise diagnoses for patients. Such advancements could translate to earlier interventions, better patient outcomes, and a deeper understanding of the disease's progression. While our study offers a compelling argument for the integration of these techniques, further research and longitudinal studies are essential to establish the longevity, reliability and widespread applicability of these findings.

**Disclosure Statement**

No potential conflict of interest was reported by the author.

Data Availability Statement

These datasets were derived from the following public domain resources: https://www.kaggle.com/datasets/tourist55/alzheimers-dataset-4-class-of-images

**Additional information**

Funding

Not applicable.

**References**

1. Javaid, S.F., Giebel, C., Khan, M.A., and Hashim, M.J. (2021). Epidemiology of Alzheimer’s disease and other dementias: Rising global burden and forecasted trends. F1000Research *10*, 425.

2. Dumurgier, J., and Sabia, S. (2020). Epidemiology of Alzheimer's disease:

latest trends. La Revue du praticien *70*, 149-151.

3. Kim, K., Kim, M.-J., Kim, D.W., Kim, S.Y., Park, S., and Park, C.B. (2020). Clinically accurate diagnosis of Alzheimer’s disease via multiplexed sensing of core biomarkers in human plasma. Nature communications *11*, 119.

4. Turner, R.S., Stubbs, T., Davies, D.A., and Albensi, B.C. (2020). Potential new approaches for diagnosis of Alzheimer's disease and related dementias. Frontiers in neurology *11*, 496.

5. Fernández Montenegro, J.M., Villarini, B., Angelopoulou, A., Kapetanios, E., Garcia-Rodriguez, J., and Argyriou, V. (2020). A survey of alzheimer’s disease early diagnosis methods for cognitive assessment. Sensors *20*, 7292.

6. Montagne, A., Barnes, S.R., Nation, D.A., Kisler, K., Toga, A.W., and Zlokovic, B.V. (2022). Imaging subtle leaks in the blood–brain barrier in the aging human brain: potential pitfalls, challenges, and possible solutions. GeroScience *44*, 1339-1351.

7. Gupta, V.B., Chitranshi, N., den Haan, J., Mirzaei, M., You, Y., Lim, J.K., Basavarajappa, D., Godinez, A., Di Angelantonio, S., and Sachdev, P. (2021). Retinal changes in Alzheimer's disease—integrated prospects of imaging, functional and molecular advances. Progress in retinal and eye research *82*, 100899.

8. Butler, E., and Mounsey, A. (2021). Structural MRI for the Early Diagnosis of Alzheimer Disease in Patients with MCI. Am Fam Physician *103*, 273-274.

9. Khagi, B., and Kwon, G.-R. (2020). 3D CNN design for the classification of

Alzheimer's disease using brain MRI and PET. IEEE Access *8*, 217830-217847.

10. Angkoso, C.V., Agustin Tjahyaningtijas, H.P., Purnama, I., and Purnomo, M.H. (2022). Multiplane Convolutional Neural Network (Mp-CNN) for Alzheimer's Disease Classification. International Journal of Intelligent Engineering & Systems *15*.

11. Woldseth, R.V., Aage, N., Bærentzen, J.A., and Sigmund, O. (2022). On the use of artificial neural networks in topology optimisation. Structural and Multidisciplinary Optimization *65*, 294.

12. Chandrasekhar, A., and Suresh, K. (2022). Approximate length scale filter in topology optimization using Fourier enhanced neural networks. Computer-Aided Design *150*, 103277.

13. Klemens, F., Schuhmann, S., Balbierer, R., Guthausen, G., Nirschl, H., Thäter, G., and Krause, M.J. (2020). Noise reduction of flow MRI measurements using a lattice Boltzmann based topology optimisation approach. Computers & Fluids *197*, 104391.

14. Lee, S., Kim, H., Lieu, Q.X., and Lee, J. (2020). CNN-based image recognition for topology optimization. Knowledge-Based Systems *198*, 105887.

15. Sasaki, H., and Igarashi, H. (2019). Topology optimization accelerated by deep learning. IEEE Transactions on Magnetics *55*, 1-5.

16. Chen, H., Jiang, Y., Ko, H., and Loew, M. (2023). A teacher–student

framework with Fourier Transform augmentation for COVID-19 infection segmentation in CT images. Biomedical Signal Processing and Control *79*, 104250.

17. Zhou, X., Qiu, S., Joshi, P.S., Xue, C., Killiany, R.J., Mian, A.Z., Chin, S.P., Au, R., and Kolachalama, V.B. (2021). Enhancing magnetic resonance imaging-driven Alzheimer’s disease classification performance using generative adversarial learning. Alzheimer's research & therapy *13*, 1-11.
18. Sathiyamoorthi, V., Ilavarasi, A., Murugeswari, K., Ahmed, S.T., Devi, B.A., and Kalipindi, M. (2021). A deep convolutional neural network based computer aided diagnosis system for the prediction of Alzheimer's disease in MRI images. Measurement *171*, 108838.
19. Yilmaz, D. (2023). Development and Evaluation of an Explainable Diagnostic AI for Alzheimer's Disease. (IEEE), pp. 1-6.
20. Marwa, E.-G., Moustafa, H.E.-D., Khalifa, F., Khater, H., and AbdElhalim, E. (2023). An MRI-based deep learning approach for accurate detection of Alzheimer’s disease. Alexandria Engineering Journal *63*, 211-221.
21. Hu, J., Wang, Y., Guo, D., Qu, Z., Sui, C., He, G., Wang, S., Chen, X., Wang, C., and Liu, X. (2023). Diagnostic performance of magnetic resonance imaging–based machine learning in Alzheimer’s disease detection: A meta-analysis. Neuroradiology *65*, 513-527.
22. Shojaei, S., Abadeh, M.S., and Momeni, Z. (2023). An evolutionary explainable deep learning approach for Alzheimer's MRI classification. Expert

Systems with Applications *220*, 119709.

23. Høilund-Carlsen, P.F., Revheim, M.-E., Alavi, A., and Barrio, J.R. (2023). FDG PET (and MRI) for monitoring immunotherapy in Alzheimer disease. Clinical Nuclear Medicine *48*, 689-691.

24. Atay, L.O., Saka, E., Akdemir, U.O., Yetim, E., Balcı, E., Arsava, E.M., and Topcuoglu, M.A. (2023). Hybrid PET/MRI with Flutemetamol and FDG in Alzheimer's Disease Clinical Continuum. Current Alzheimer Research.

25. Erickson, P., Simrén, J., Brum, W.S., Ennis, G.E., Kollmorgen, G., Suridjan, I., Langhough, R., Jonaitis, E.M., Van Hulle, C.A., and Betthauser, T.J. (2023). Prevalence and Clinical Implications of a β-Amyloid–Negative, Tau-Positive Cerebrospinal Fluid Biomarker Profile in Alzheimer Disease. JAMA neurology *80*, 969-979.

26. Gao, F., Dai, L., Wang, Q., Liu, C., Deng, K., Cheng, Z., Lv, X., Wu, Y., Zhang, Z., and Tao, Q. (2023). Blood-based biomarkers for Alzheimer’s disease: a multicenter-based cross-sectional and longitudinal study in China. Science Bulletin *68*, 1800-1808.

27. Malek-Ahmadi, M., Duff, K., Chen, K., Su, Y., King, J.B., Koppelmans, V., and Schaefer, S.Y. (2023). Volumetric regional MRI and neuropsychological predictors of motor task variability in cognitively unimpaired, Mild Cognitive Impairment, and probable Alzheimer's disease older adults. Experimental Gerontology *173*, 112087.

28. Merlin, G., and Pattabiraman, V. (2023). Prediction and Feature Extraction

Techniques used for Classification of Alzheimer’s Disease in its Early Stage using MRI: A Review. (Atlantis Press), pp. 94-100.

**Figure legends**

Figure 1. Title: Flowchart for Image Improvement and Alzheimer's Disease Classification Using CNN on MRI Images. The flowchart encompasses two main processes; Image Improvement and Image Classification. The Image Improvement process begins with "Convolution", followed by "Fourier Transform" to transition to the frequency domain, then "Frequency Domain Multiplication" for filtering, defining a "Low-pass Filter" and "Subtracting the Low-pass Component" for emphasizing details, "Applying the High-pass Filter" for edge enhancement, transitioning back to spatial domain with "Inverse Fourier Transform of the Result", "Normalization" for data consistency, and finally "Discrete Implementation" for practical application. On the other hand, the Image Classification process kicks off with "Data Acquisition" of MRI images, "Data Preprocessing" for structuring the dataset, defining the "Model Architecture (CNN)" for classification, "Model Compilation" with suitable metrics, "Model Training" using the training dataset, "Model Evaluation" on the testing dataset to gauge performance, and "Result Interpretation" to assess the predictive accuracy and model's efficacy in classifying Alzheimer's disease stages.

Figure 2. MRI (Magnetic Resonance Imaging) scans of human brains. A, raw images. B, the images treated via topological optimization.

Figure 3. Metrics evaluating the performance of a convolutional neural network (CNN) over 100 epochs. A, the first graph is titled "Model ACC" and plots the model's accuracy, with two lines representing "train" (in blue) and "val" (in orange). The second graph, "Model AUC", showcases the area under the receiver operating characteristic curve for both the training and validation sets, again distinguished by the "train" (in blue) and "val" (in orange) legends. The third graph, "Model Loss", tracks the model's loss across epochs, with the "train" (in blue) curve representing training loss and the "val" (in orange) curve indicating validation loss. B, the performance after topological optimization

Figure 4. Confusion Matrix for Alzheimer's Disease Diagnosis via CNN Classification. A, this figure presents a matrix juxtaposing the actual diagnostic categories ("Truth") against the convolutional neural network's predictions. The color gradient, ranging from light to dark green, represents the count of instances, with darker shades indicating higher counts. Within each cell, numbers provide the exact

count of CT images classified. Diagonal cells indicate correct predictions, while off-diagonal cells capture misclassifications.

Figure 1

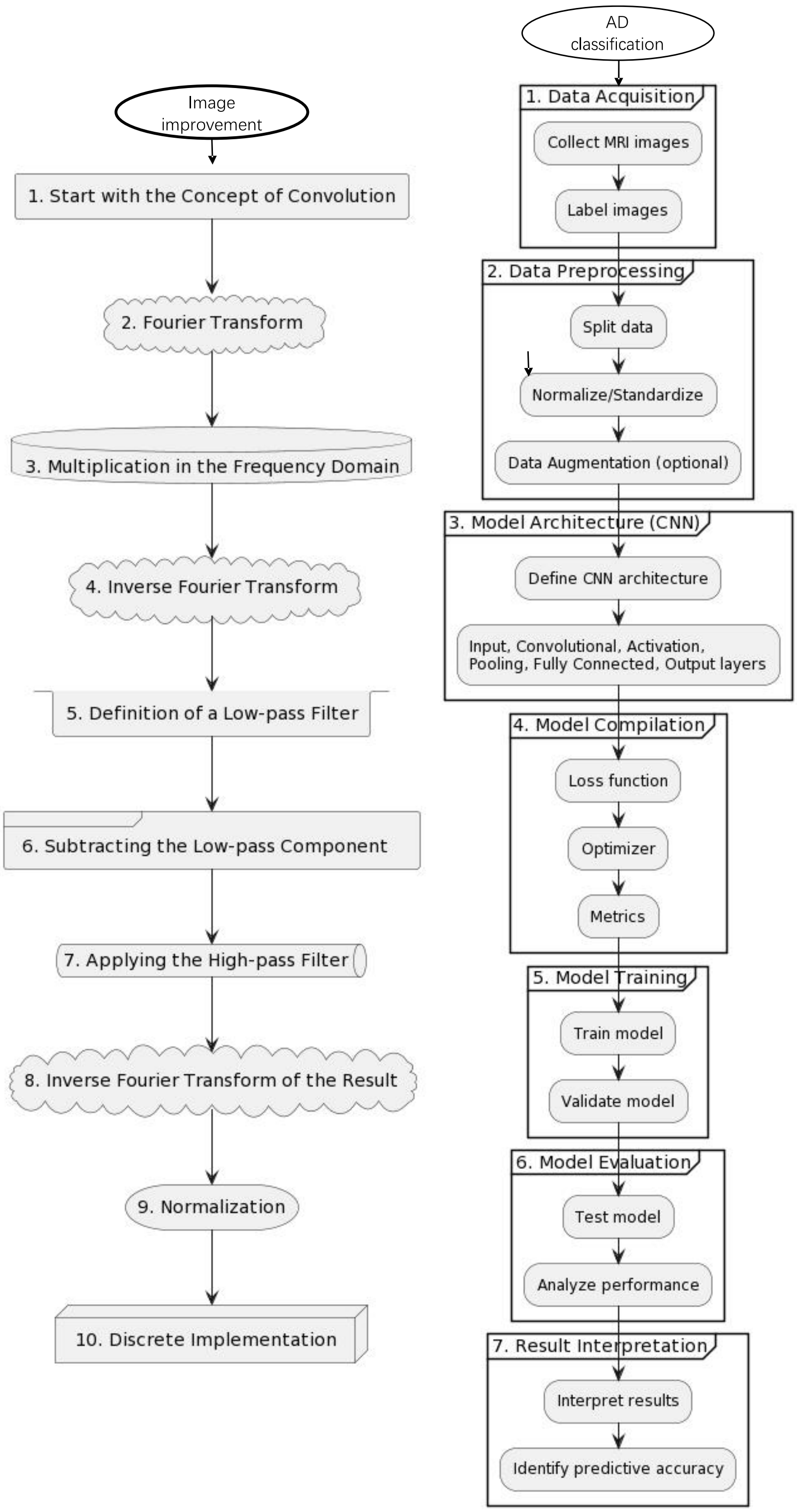

Image improvement
1. Start with the Concept of Convolution
2. Fourier Transform
3. Multiplication in the Frequency Domain
4. Inverse Fourier Transform
5. Definition of a Low-pass Filter
6. Subtracting the Low-pass Component
7. Applying the High-pass Filter
8. Inverse Fourier Transform of the Result
9. Normalization
10. Discrete Implementation
AD classification
1. Data Acquisition
Collect MRI images
Label images
2. Data Preprocessing
Split data
Normalize/Standardize
Data Augmentation (optional)
3. Model Architecture (CNN)
Define CNN architecture
Input, Convolutional, Activation, Pooling, Fully Connected, Output layers
4. Model Compilation
Loss function
Optimizer
Metrics
5. Model Training
Train model
Validate model
6. Model Evaluation
Test model
Analyze performance
7. Result Interpretation
Interpret results
Identify predictive accuracy

Figure 2A

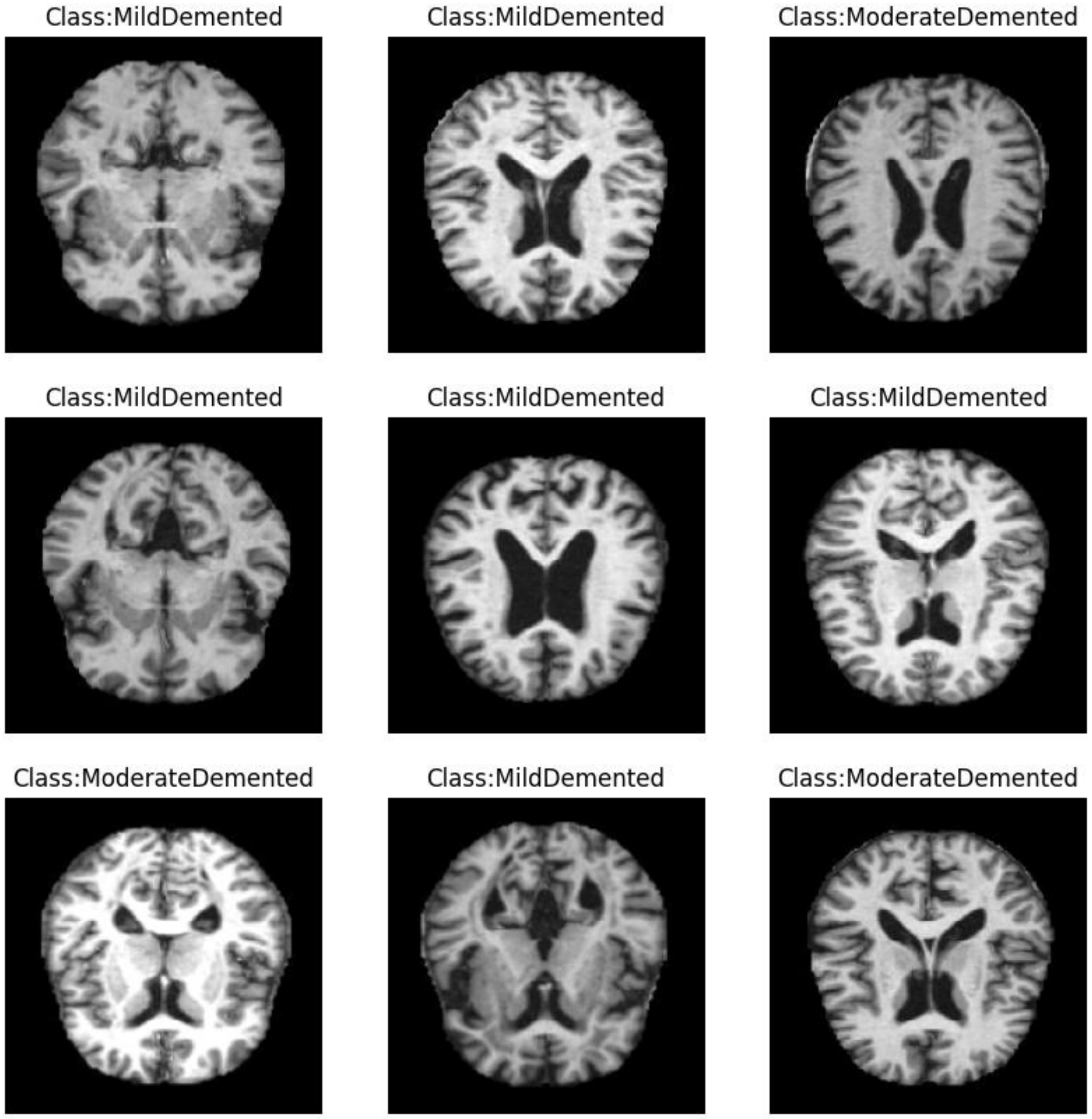

Figure 2B

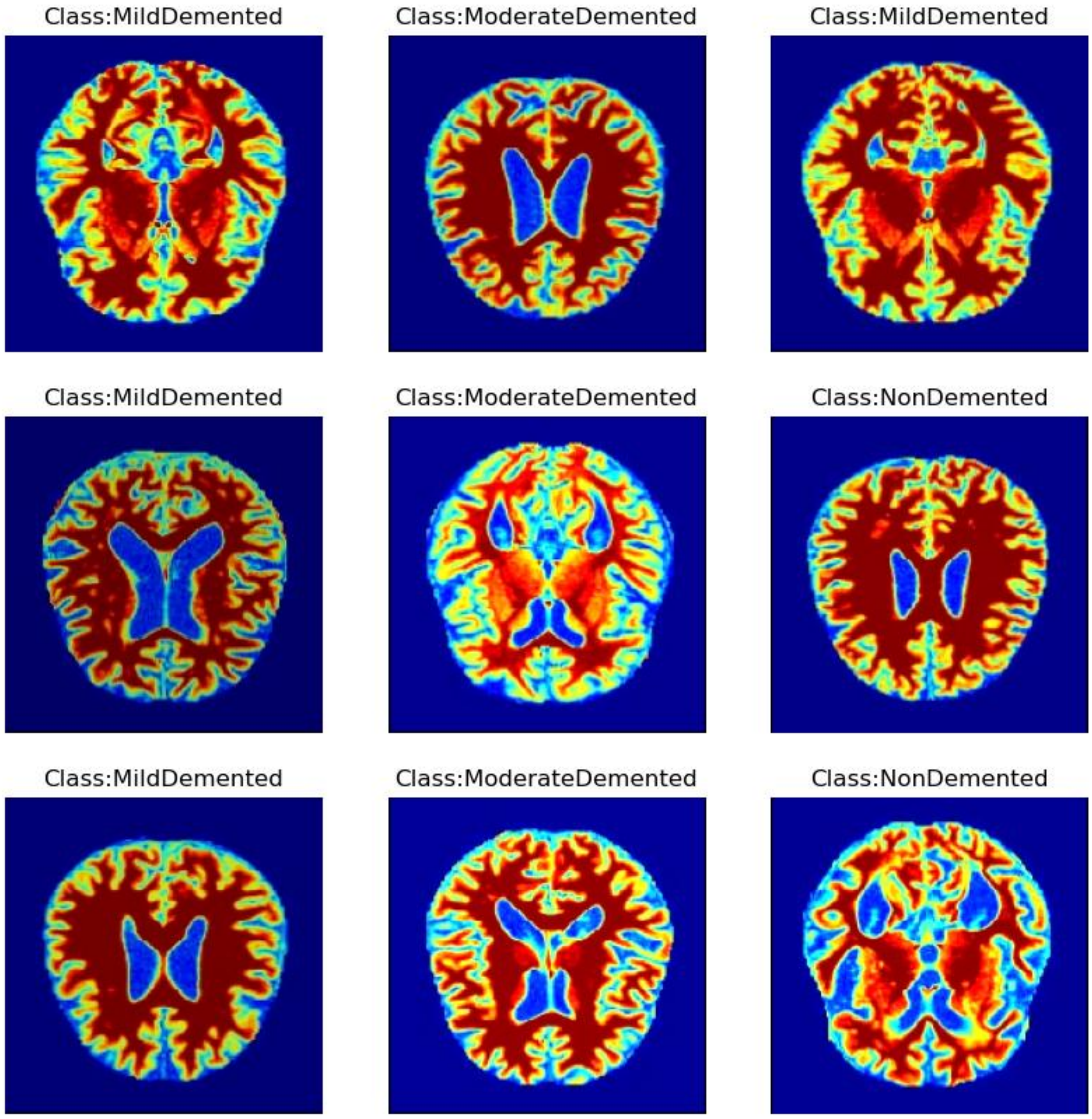

Figure 3A

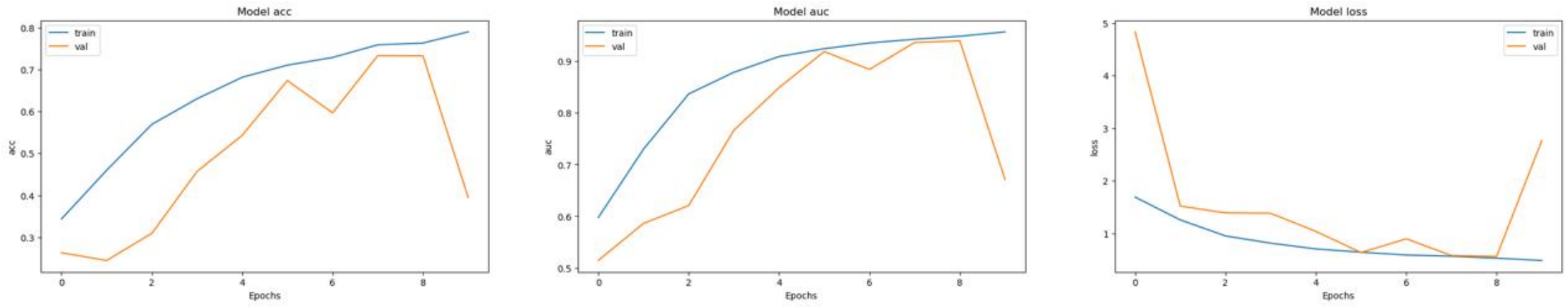

3B

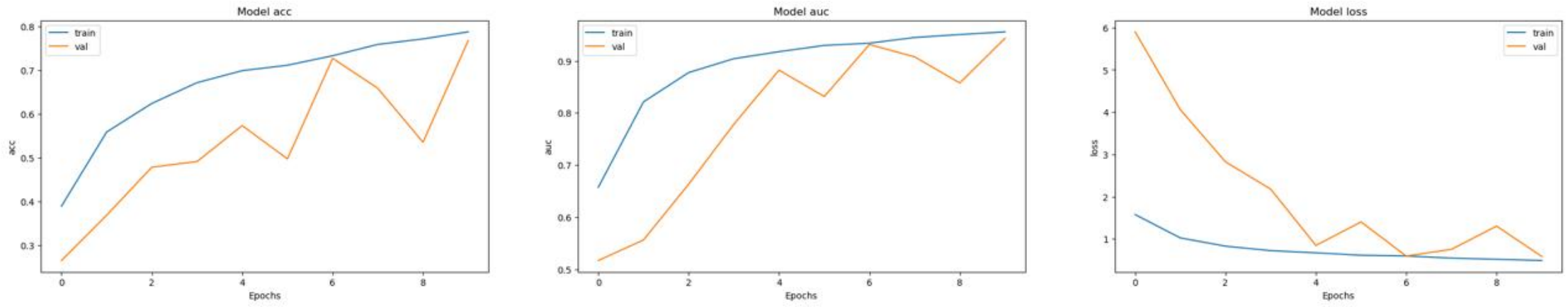

3C

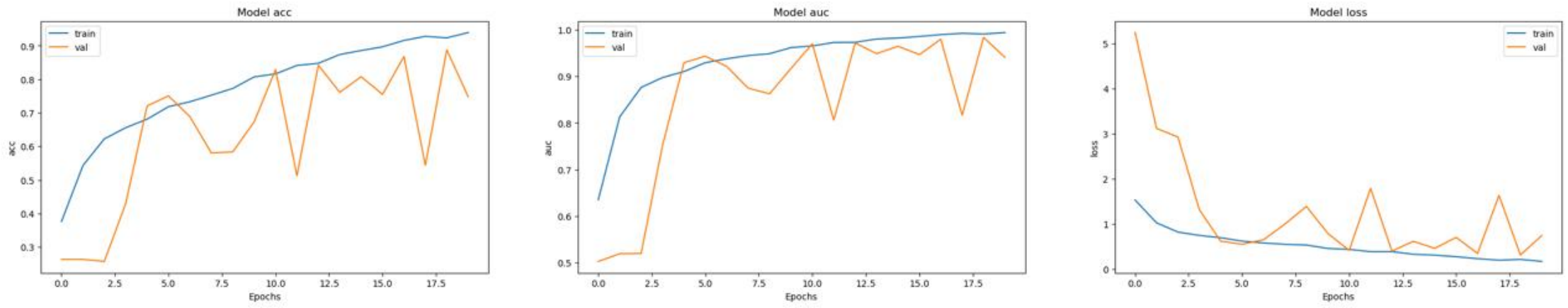

3D

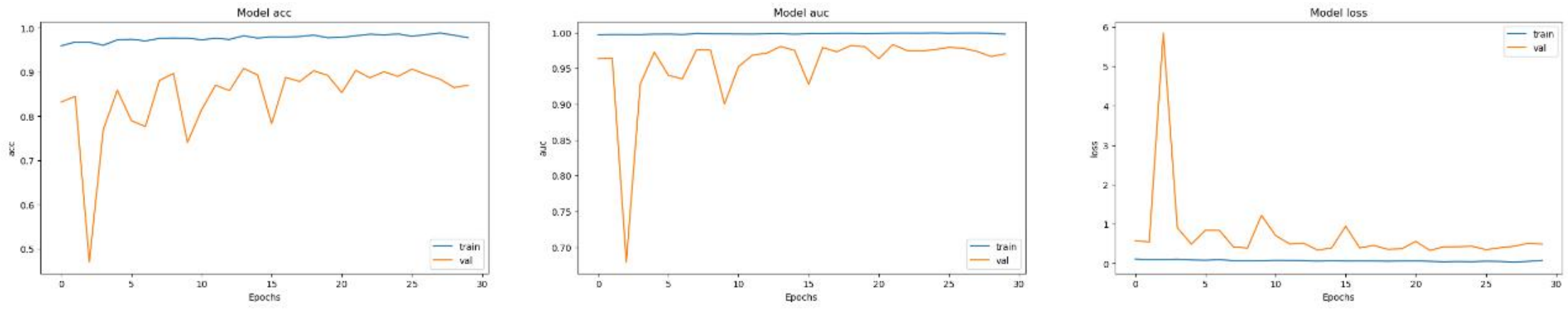

Figure 4A

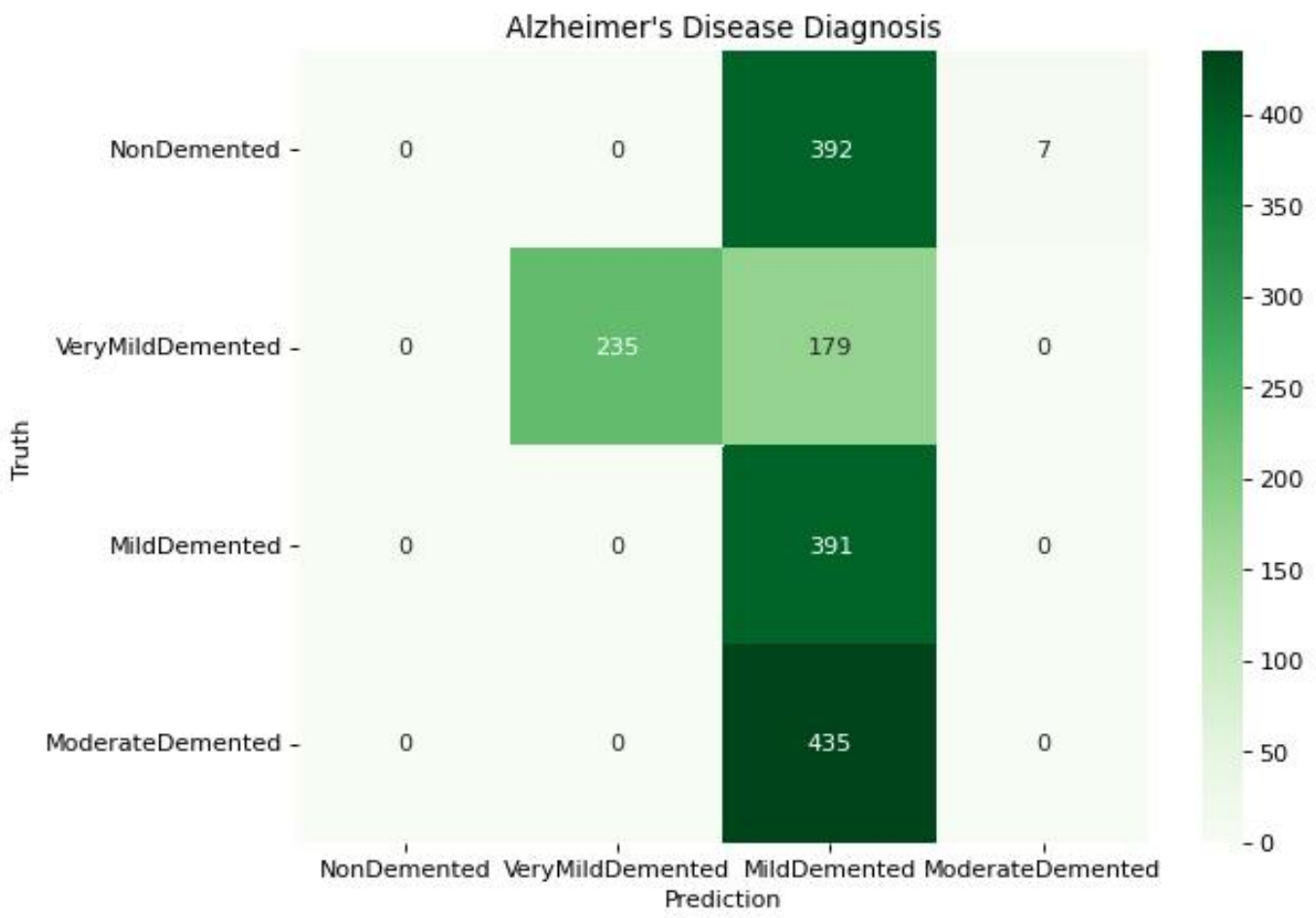

4B

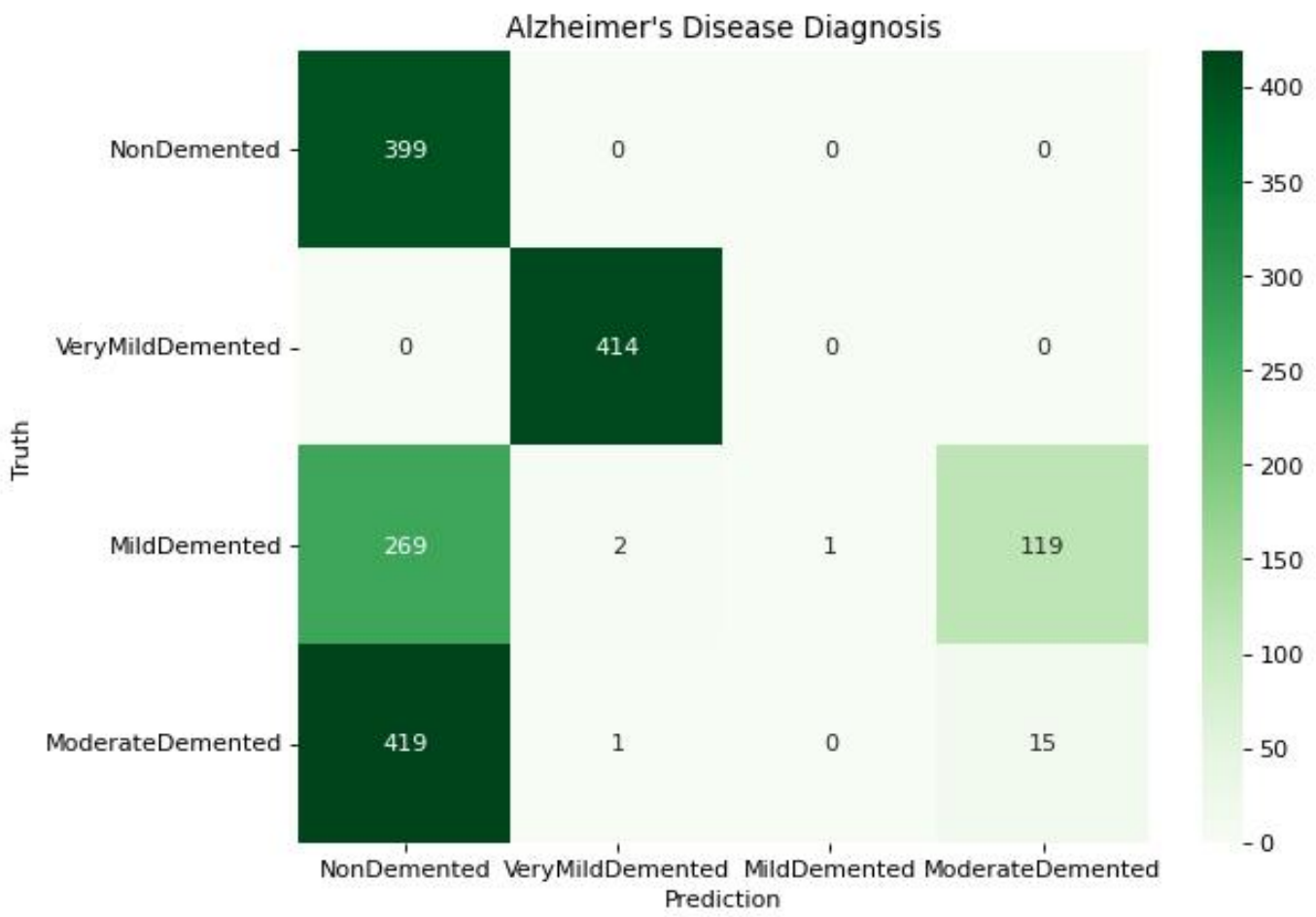

4C

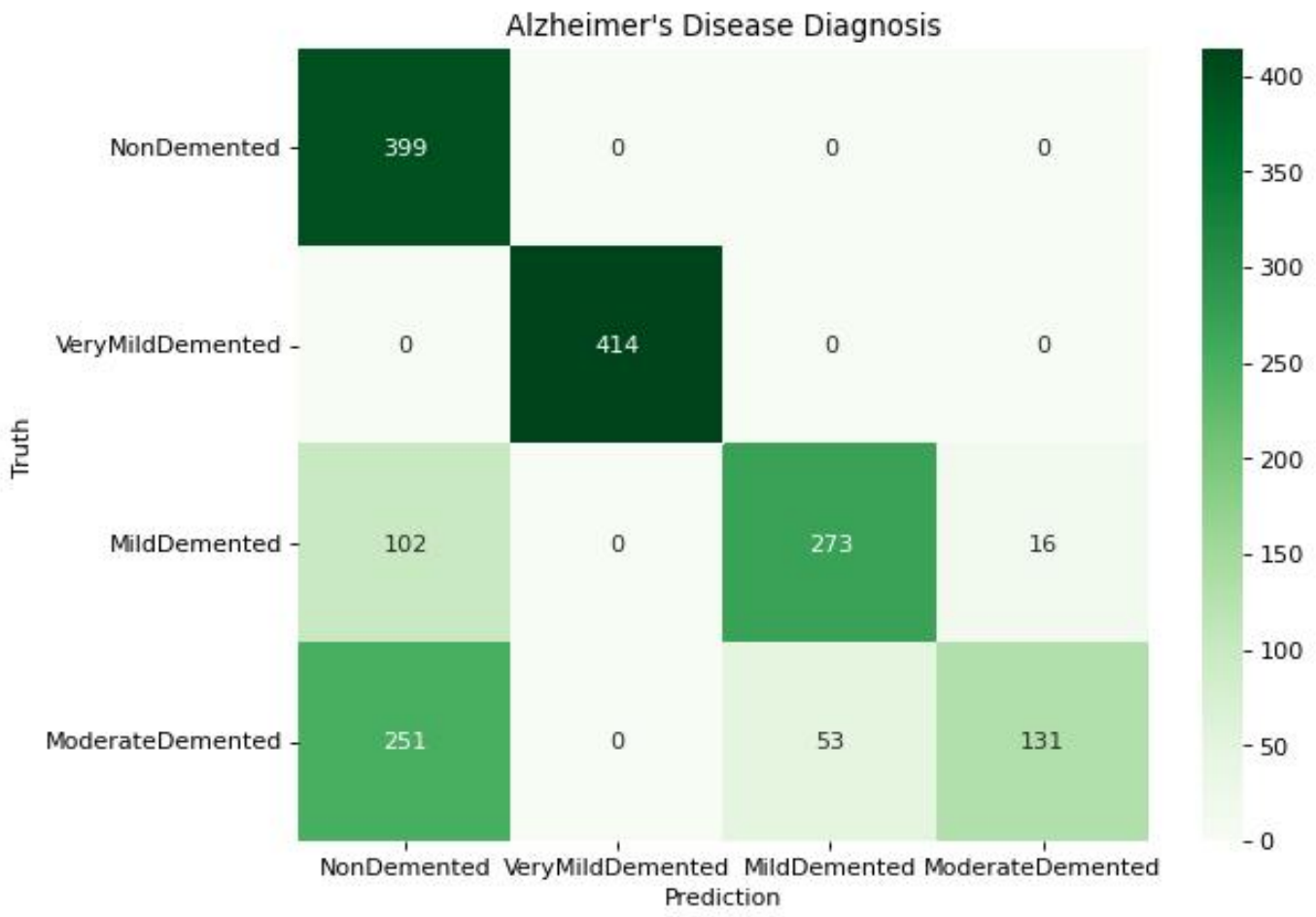

4D

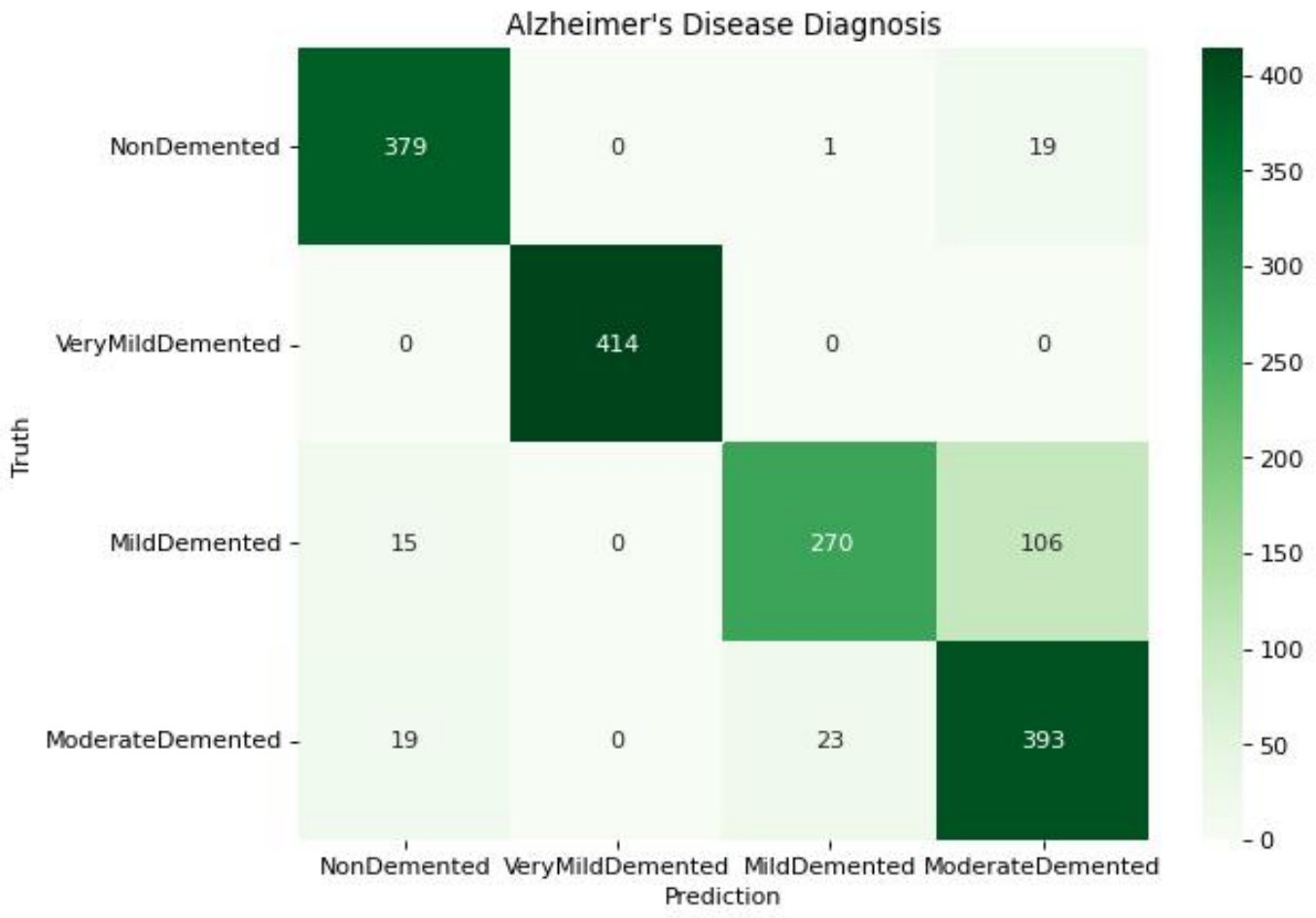